\documentclass[manuscript,screen,natbib=false]{acmart}

\AtBeginDocument{%
  }

\setcopyright{acmlicensed}
\copyrightyear{2018}
\acmYear{2018}
\acmDOI{XXXXXXX.XXXXXXX}
\acmConference[Conference acronym 'XX]{Make sure to enter the correct
  conference title from your rights confirmation email}{June 03--05,
  2018}{Woodstock, NY}
\acmISBN{978-1-4503-XXXX-X/2018/06}

\RequirePackage[
  datamodel=acmdatamodel,
  style=acmnumeric,
  ]{biblatex}

\usepackage[table]{xcolor}
\usepackage{booktabs}
\usepackage{multirow}
\usepackage{bigdelim}
\usepackage{csquotes}
\usepackage{subcaption}

\begin{document}

\title[Impact Matters!]{Impact Matters! An Audit Method to Evaluate AI Projects and their Impact for Sustainability and Public Interest}


\author{Theresa Züger}
\affiliation{%
 \institution{Alexander von Humboldt Institute for Internet and Society}
\country{Germany}
}

\author{Laura State}
\affiliation{%
  \institution{Alexander von Humboldt Institute for Internet and Society}
\country{Germany}
}

\author{Lena Winter}
\affiliation{%
\institution{Alexander von Humboldt Institute for Internet and Society}
\country{Germany}
}

\renewcommand{\shortauthors}{Züger et al.}

\begin{abstract}
The overall rapid increase of artificial intelligence (AI) use is linked to various initiatives that propose AI `for good'. However, there is a lack of transparency in the goals of such projects, as well as a missing evaluation of their actual impacts on society and the planet.
We close this gap by proposing public interest and sustainability as a regulatory dual-concept, together creating the necessary framework for a just and sustainable development that can be operationalized and utilized for the assessment of AI systems.
Based on this framework, and building on existing work in auditing, we introduce the Impact-AI-method, a qualitative audit method to evaluate concrete AI projects with respect to public interest and sustainability. The interview-based method captures a project's governance structure, its theory of change, AI model and data characteristics, and social, environmental, and economic impacts.
We also propose a catalog of assessment criteria to rate the outcome of the audit as well as to create an accessible output that can be debated broadly by civil society.
The Impact-AI-method, developed in a transdisciplinary research setting together with NGOs and a multi-stakeholder research council, is intended as a reusable blueprint that both informs public debate about AI `for good' claims and supports the creation of transparency of AI systems that purport to contribute to a just and sustainable development. 
\end{abstract}





\maketitle

\section{Introduction}

Within the overall rapid rise of artificial intelligence (AI) use in the last decade, the narrative of AI systems being utilized `for good' in specific use cases has manifested. 
This includes AI applications for sustainability, such as enhancing wildlife protection by tracking endangered species or collecting litter from the ocean floor, to socially oriented use cases such as supporting equality in online discussions or augmenting the work of mediators in national conflict.\footnote{Such projects are: \url{https://www.zambacloud.com};  \url{https://seaclear-project.eu/}; \url{https://stanforddeliberate.org/}; \url{https://negotiatecop.org}} 
Numerous state and private initiatives, as well as research and civil society institutions, have committed to supporting, researching, developing, and financing the goal of using AI for social and sustainable benefit.\footnote{Such initiatives are: \url{https://aiforgood.itu.int/; https://www.sdgaicompendium.org/}; \url{https://www.civic-coding.de/en/home}; \url{https://www.microsoft.com/en-us/research/group/ai-for-good-research-lab/}; \url{https://de.mi4people.org/}; \url{https://www.dagstuhl.de/en/seminars/seminar-calendar/seminar-details/26021}}
This commitment turns the pursuit of applying AI `for good' not only into a narrative but also into a research and development program \cite{aula_stepping_2023} which is enacted in a dispersed and uncoordinated manner by different actors, seemingly having one purpose in common: to serve the improvement of society. At the same time, the ethical implications of these goals are rarely made transparent or discussed by the owners of these projects publicly.
Therefore, their intentions suffer from a vagueness of what it means to serve `the good' or even the Sustainable Development Goals--often seen as a proxy of collective global goals that lead to a more sustainable and just world \cite{Patton_2023}.
Additionally, as critical scholars have pointed out in the field of AI `for good', underlying problems are not taken into account sufficiently, such as the exploitation of labor of data workers and natural resources \cite{birhane_false_2025}. Even more, many of the implemented uses cases of AI `for good' are not questioned and evaluated in regards to their actual impacts.

As a remedy for this lack of transparency and evaluation, we developed a comprehensive audit methodology, the {Impact-AI-method}, which allows to systematically analyze and assess concrete use cases of AI systems in regards to their impacts.
We thereby build on sustainability and public interest as theoretical concepts and \enquote{regulatory double-idea} \cite{weidner_gemeinwohl_nodate}. Their combination represents a suitable frame for understanding the most relevant objectives and tensions to enable the necessary negotiation process for a just and sustainable development.
Additionally, other than ethical frameworks, both concepts are well-established with practical implications: sustainability and public interest have been operationalized by existing research and societal practices, which makes them more concrete than guiding values or principles \cite{rohde_nachhaltigkeitskriterien_2021, zuger_ai_2023, weidner_gemeinwohl_nodate, feintuck_public_2004,harrach_transformation_2023}. Thus, they can be investigated and to a certain degree assessed, to allow for public transparency, accountability, and debate. 

With this work, we therefore aim to answer the following research question: \textit{How can the impact of specific AI use cases be analyzed, building on the concepts of sustainability and public interest?}
The main contributions of this paper are the following: 
\begin{itemize}
    \item [1.] We are combining and operationalizing the concepts of public interest and sustainability as a \textbf{theoretical framework} for the assessment of AI systems to cover the most relevant objectives and tensions underlying just and sustainable development. 
    \item [2.] Building on this framework, we propose the \textbf{Impact-AI-method}, an audit method to understand and create evidence for the impacts of different AI projects for concrete societal goals, and support transparency on these projects.
    Part of the method is an assessment catalog to rate the audit results. By translating the results into an easily understandable overview, we make them accessible to a wider audience.
\end{itemize}

\paragraph{Scope}
We developed the {Impact-AI-method} with a focus on projects that aim to make a contribution to the public interest or sustainability. 
However, our method is also applicable to other AI projects that do not make an active claim to have these societal beneficial objectives. 
Also, the Impact-AI-method can be applied to various AI models and is not specific to a domain such as computer vision or natural language processing.
These two generalizations are made possible through the general approach of the Impact-AI-method, and certain tailoring steps that need to be followed when preparing the interviews for a specific AI project.

Additionally, the audit method is first of all designed for researchers and stakeholders such as NGOs and public bodies and does not provide any standardization, labels, or legal compliance checks. 

\paragraph*{Structure of the paper}
In Section~\ref{sec:theoretical_background}, we provide the theoretical background of the two key concepts behind our method: public interest and sustainability. 
Section~\ref{sec:related_work} discusses related work on auditing, documentation and ethics guidelines, Section~\ref{sec:method_development} explains the development of the {Impact-AI-method}, and Section~\ref{sec:impact_ai_method} introduces the method.
In Section~\ref{sec:discussion}, central controversies, namely the relationship between our method and AI compliance audits, the comparability of audit results, and the question of impact of the audit method itself, are discussed. The paper ends by reflecting on our contributions and describing the pathways for necessary and beneficial future work in Section~\ref{sec:conclusion_outlook}.

\section{Theoretical Background}
\label{sec:theoretical_background}

In this section, we discuss the concepts of `public interest' and `sustainability' that are core to our framework, as well as how these two concepts relate to each other.

\subsection{How Do We Understand the Public Interest?}

The public interest--a term used sometimes interchangeably with the `common good'\footnote {While some theorists use the public interest and the common good as synonyms, others see differences. Douglass \cite{douglass_common_1980} understands common good as something which affects everyone, is shared, is more important than individual good, and a higher political authority determines what is included in the common good. In contrast, he interprets public interest as what is really good for all the people. In a democratic society, this would mean: what is really good for people as a whole and interpreted by the people.}--is a widely discussed concept with a long history as \enquote{a highly contingent, normative term in politics, law and philosophy} \cite{von_der_pfordten_2008}. It is deeply rooted within democratic values and is often seen as a goal of democratic governance overall \cite{bozeman_public_2007}. However, there is no universally shared definition of the public interest and also how to achieve it is contested, which is not surprising due to its highly normative character. Notwithstanding, the concept has proven its practical value in many contexts, such as law cases and public administration \cite{munkler_gemeinwohl_2002}. In the last two decades, public interest has experienced a renaissance in recent political debates \cite{Offe_2012} and an \enquote{active incarnation in public policy and management} \cite{bozeman_public_2007}. This understanding highlights the public as a pluralistic body within which a participatory process of self-determination needs to take place in order to realize the notion of the public interest. The `public' gives reference to a specific community which is the body of people concerned. Such a `body' might refer to very different sizes, ranging from a regional community, to a country, to supranational community such as the European Union, or even more global publics. 
Public interest theories, as articulated by Barry Bozeman \cite{bozeman_public_2007}, Virginia Held \cite{held_1970} or Ernst Fraenkel \cite{fraenkel_1979}, and many other theorists, discuss not how individual and private interest guide the actions of a society, but rather how collectively negotiated interests of a public can be realized. Many theorists agree that public interest(s) need to be defined for each societal issue individually, combining a normative and proceduralist approach, which is also described as a hybrid approach to public interest \cite{hiebaum_hybride_2020}. Aside from commonly shared norms, such as justice or other foundational norms of democracy, these theories often stress the importance of participation and an equitable deliberative process as a constituent element of the public interest and its realization in practice \cite{bozeman_public_2007}. Bozeman argues that the pursuit of the public interest \enquote{is a matter of using open minds and sound, fair procedures to move ever closer to the ideal that is revealed during the process and, in part, by the process} \cite{bozeman_public_2007}. Similarly, Weidner \cite{weidner_gemeinwohl_nodate}  sees public interest as the result of a fair, pluralistic balancing of interests, achieved through recognized procedures and compromise--which is a view often shared and followed by modern democratic constitutions. Public interest in pluralistic, modern societies is not a status quo, a label with which `right' actions are rewarded. It is never a static social reality, but a `regulatory idea' \cite{fraenkel_1960}, that guides governance and societal problem solving.

According to this theoretical understanding, we do not think of public interest as a status that can be verified through our method. Rather, our approach aims to make the relevant procedural logic and criteria aiming for the public interest--such as participatory design, transparency, and openness for validation--and the implementation of these criteria visible and debatable for the academic and public discourse. 

\paragraph{Public interest and AI}
In previous work, scholars have developed criteria for public interest AI \cite{zuger_ai_2023,zuger_civic_2022}, which translate principles of public interest theory and expert knowledge from the field of public interest tech into actionable points for the development and reflection of AI projects.
These include 1) public justification: giving a publicly transparent argument, why AI is implemented, why this solution serves a public interest, and why it outperforms alternatives; 2) equity: the assurance that the AI system does not only discriminate, but serves equity and equality in their application purpose, but also by design; 3) deliberation and participatory design: providing not only transparency for users but also allowing and inviting affected stakeholders to influence the design and governance of the project in a meaningful way; 4) technical safeguards: technical standards and safeguards of the system including the quality of data, the system accuracy, the data privacy and the security of system's infrastructure; 5) openness for validation: meaning the possibility for third parties to validate the functions of the system according to its intended impacts. 
All these criteria are considered and included in the {Impact-AI-method}. More details on how the concept of public interest is part of our method can be found in Section~\ref{sec:impact_ai_method}.

\subsection{How Do We Understand Sustainability?}
\label{sec:public_interest}

The first internationally recognized and commonly referred to definition of sustainability stems from the report \textit{Our common future} from 1987 \cite{WCED}. The report published by the \textit{World Commission on Environment and Development} (WCED), better known as \textit{Brundtland Report}, formulates sustainability as follows: \enquote{Humanity has the ability to make development sustainable to ensure that it meets the needs of the present without compromising the ability of future generations to meet their own needs.} \cite{WCED}. It frames sustainability as an ongoing development and continuous process, encompassing both normative objectives and procedural elements \cite{spindler_history_2013}. Also, it notably integrates environmental concerns with global development objectives and adopts a holistic, long-term perspective \cite{kopfmuller_auf_2007}. Integral part of this understanding of sustainability are both intergenerational justice, i.e., responsibility toward future generations, as well as intragenerational justice, i.e., addressing the equitable distribution of resources among current populations \cite{mensah_sustainable_2019}. In summary, this definition extends beyond environmental concerns, incorporating social and economic dimensions with the overarching aim of securing equal or improved living conditions for present and future generations.

Building on this foundation, the 2015 \textit{Agenda for Sustainable Development} and the 17 Sustainable Development Goals (SDGs) \cite{2030_Agenda_for_sustainable_development} represent a holistic set of goals that integrates the three dimensions of social, environmental, and economic sustainability. Recognizing that social and economic systems cannot be delineated from the environmental dimension, various models have been developed to conceptualize sustainability. Most emphasize the three dimensions--economic, social, and environmental--but differ in how these are related.
Some models place all three dimensions on equal footing, suggesting that sustainability requires balanced consideration of each, see, for instance, \cite{Elkington_1994}. In contrast, hierarchical models argue that the natural environment forms the foundation upon which society and the economy depend and should therefore take precedence in sustainability discourse \cite{spindler_history_2013}.
Although we acknowledge the ongoing challenges in defining sustainability, we support a multidimensional sustainability understanding and the interrelatedness and synergies between its different dimensions. We opt for a priority model of sustainability as it is presented in the so-called \textit{Wedding Cake} model which visualizes and highlights an hierarchical order of the different SDGs \cite{stockholm_resilience_2016}. The model gives priority to the environmental dimension, followed by social goals and lastly economic ones. It sees the biosphere as a precondition for sustainable development and as foundation on which social and economic goals depend on \cite{folke_social-ecological_2016}. This understanding of a hierarchy is underpinned by the concept of the planetary boundaries, introduced by Rockström et al. \cite{Rockström_2009}. It underscores the environmental limits within which human activity must be constrained to ensure the planet’s viability for current and future generations and identifies nine critical Earth system processes that define a \enquote{safe operating space} for humanity. Transgressing these boundaries increases the risk of large-scale environmental degradation and destabilization of the ecosystems leading to systemic disruptions to human and environmental health. 

\paragraph{Sustainability and AI}
\label{sec:sustainability_and_ai}

Sustainability and AI are connected by two contradictory narratives at first glance: 
on the one hand, with the increased use of large language models (LLMs) such as the GPT-models, AI is increasingly scrutinized for their use of natural resources and for causing additional greenhouse gas emissions, thus accelerating climate change.
Scholars have investigated their power consumption and related emissions during training and inference \cite{luccioni_power_2024,DBLP:journals/jmlr/LuccioniVL23}, their water consumption \cite{li_making_2025-1}, the land use of data centers \cite{crawford_2021}, and their use of raw materials \cite{crawford_2021}, among others.
There are also several tools that help to measure energy consumption and estimate the associated footprint, with the general objective of reducing the environmental damage that AI models cause.\footnote{Examples are Code Carbon \url{https://codecarbon.io/}, Carbon Tracker \url{https://carbontracker.info/} and ML CO2 Impact \url{https://mlco2.github.io/impact/}}
On the other hand, AI systems are seen as technologies that can support efforts in the fight against climate change \cite{rolnick_tackling_2023,kaack_aligning_2022}. Specific examples of such AI systems are their use in the optimization of industrial cooling processes to save resources or in monitoring peatlands, a natural landscape that is essential to saving carbon \cite{rolnick_tackling_2023}.
Both, the increasing need of AI for natural resources and the potential examples of AI to support efforts for environmental sustainability exist. However, they can hardly be weighed against each other with the aim of calculating the remaining benefit or burden of AI usage on a global scale. 
Instead, systems development, benefits, and costs must be analyzed case by case and in regards to all levels of its societal and public interest oriented impacts as well as a multidimensional understanding of sustainability. 
%
Details on how the concept of sustainability is part of our method can be found in Section~\ref{sec:impact_ai_method}.

\subsection{Public Interest and Sustainability: a Productive Tension}

Our theoretical framework builds on the combination of the concepts of public interest and sustainability. In agreement with Weidner \cite{weidner_gemeinwohl_nodate} we see their combination as necessary for future decision-making enabling sustainable and just development. 
Weidner describes a productive tension between both concepts, which means that they enrich each other mutually from a theoretical perspective but also in practical application.

As Weidner, we believe that both concepts are to be evaluated on an equal and not hierarchical level. Each concept on its own appears \enquote{unsuitable in political and social practice for solving pressing global problems and shaping a global civil society} \cite{weidner_gemeinwohl_nodate}. Therefore, as Weidner argues, a `regulatory dual concept' is essential: in the spirit of complementarity, it must incorporate the seemingly irresolvable (dialectical) tension between the two concepts and guide a political learning and decision-making process. Weidner concludes that the combined consideration of public interest and sustainability is \enquote{unavoidable, since under current global development trends there can be no public interest without sustainability, but also no sustainability without public interest}\footnote{Translated by the authors.} \cite{weidner_gemeinwohl_nodate}.

On the one hand, sustainability and public interest have a strong overlap, for example, both aiming for collective well-being, equity, and the conditions for human survival. 
However, the two concepts also differ: while public interest tends to provide arguments for a specific, rather local community and interests that have a meaning at a given time even though a long-term perspective might be relevant, sustainability inherently builds on a perspective that aims for justice between current and future generations and tends towards a global viewpoint. These different perspectives require a necessary process of global and local negotiation to find practical solutions to partly conflicting objectives and interests. 
Furthermore, sustainability as a multidimensional concept is already well spelled out in societal practices but has normative weaknesses in its internal trade-offs \cite{kroll_sustainable_2019}, for example, when prosperity is served to the expanse of societal benefits. Conversely, the public interest perspective centers the question of purposeful action and introduces emphasis on the procedural quality of how this purpose is achieved by requiring a participatory and deliberative approach as negotiation of interests. 
This negotiation is relevant to many issues connected to just and sustainable~development.

This also holds for AI systems and understanding when and how these systems support such a just and sustainable development.
Therefore, we use the regulatory double idea of public interest and sustainability as theoretical framework for the Impact-AI-method. Or, said differently, the Impact-AI-method is a concrete application of the regulatory dual concept of public interest and sustainability to assess AI systems and their impact for societal goals.


\section{Related Work: Auditing, Documentation, and Ethics Guidelines}
\label{sec:related_work}

In this work, we understand an audit as \enquote{{any independent assessment of an identified audit target via an evaluation of articulated expectations with the implicit or explicit objective of accountability}}~\cite{DBLP:conf/satml/BirhaneSOVR24}.
An AI audit can be conducted {internally} or {externally} to the organization deploying the AI system \cite{DBLP:conf/satml/BirhaneSOVR24,li_making_2025}. Furthermore, it can take place before the AI system is deployed ({ex ante}), during design and re-deployment ({in media res}), or when the system is already deployed ({ex post})~\cite{DBLP:conf/satml/BirhaneSOVR24}.
More specifically, such an audit can be conducted by {academic} or {non-academic} actors~\cite{DBLP:conf/satml/BirhaneSOVR24}. A simple taxonomy of audits defines five different types \cite{li_making_2025}: 1) governance/compliance audits, which check for alignment with government policies; 2) ethics-based audits, examining AI systems for ethical principles and values; 3) empirical audits, relying on input-output pairs to measure what the algorithm does; 4) technical audits, checking for robustness and \enquote{working as intended} AI systems; and 5) data audits and review, in which data collection and associated processes are checked by governance bodies.

The {Impact-AI-method} is a method designed for {external}, {ex post} auditing, and can be broadly classified as an {ethics-based} audit, although it also contains elements of {empirical} and {technical} audits.
However, it is neither a compliance nor a data audit and review. This does not mean that the legal framework in which the AI system is operating does not matter; we certainly do take this into consideration during audit preparation. However, we go beyond compliance, as our method aims to understand the purpose behind AI projects and its impacts on all levels of sustainability and public~interest.

Auditing an AI system often focuses on societal impact of such a system such as fairness and transparency\footnote{For example, this survey paper \cite{ojewale_towards_2025} is listing various AI auditing tools that cover aspects such as fairness, explainability and privacy of an AI system but no environmental aspects.} rather than impacts of the same system on the {environment}, an aspect that becomes increasingly important--see also Section~\ref{sec:public_interest}, Paragraph \enquote{Sustainability and AI}.
Approaches that bring both sides together often arrive under the concept of `multidimensional sustainability', i.e., acknowledging its social, environmental, and economic dimension.
As of now, we know only one practical approach that is using such a comprehensive understanding to assess AI systems:
Rohde et al. are putting forward 19 assessment criteria of AI systems, covering the three sustainability dimensions. Each criterion comes with an operationalization (the 
indicators) and is mapped to the life-cycle of an AI system~\cite{rohde_broadening_2024}.
In addition to the academic work of Rohde et al., the associated project \textit{SustAIn}\footnote{\url{https://sustain.algorithmwatch.org/en/about-us/about-the-sustain-project/}} provides a self-assessment that can be utilized by AI system providers and gives a first idea of how their systems score w.r.t. the proposed sustainability criteria.
This work is an important building block for the {Impact-AI-method}: it served as one of the key papers on which we relied when initially drafting our method.
However, our work is also different: in contrast to Rohde et al.~\cite{rohde_broadening_2024}, we do not only focus on supervised AI models but designed a method that can be tailored to different types of AI models. We also go beyond indicators that allow for relatively simple \enquote{yes-no} answers, proving an approach that draws mostly on qualitative methods. 
Last but not least important, our method additionally incorporates the concept of public interest which we deem essential for sustainable development, {as it emphasizes the participatory approach necessary to achieve such a common interest.}

Closely related to auditing and essential for the development of the {Impact-AI-method} is the work on {documentation}. 
The first key contribution is the model card framework~\cite{DBLP:conf/fat/MitchellWZBVHSR19}. This card is a short description of the most relevant aspects of an AI model. The relevant sections of such a card that translated into the Impact-AI-method are model details, the intended use of an AI model, factors that are relevant for its performance, evaluation metrics, ethical considerations, as well as caveats and recommendations.
Equally relevant to the Impact-AI-method is the datasheet framework~\cite{DBLP:journals/cacm/GebruMVVWDC21}. Similarly to the model cards, a datasheet summarizes relevant aspects of a dataset. They cover the motivation behind dataset creation, its composition and collection process, dataset use, and maintenance. The authors formulate questions for each of these aspects that we used as inspiration for our method.
There is also a growing body of work on how energy use and other aspects of environmental sustainability can be documented for an AI model~\cite{DBLP:conf/pkdd/0001JMM22,DBLP:journals/corr/abs-2002-05651}. Although individual aspects of such documentation strategies were certainly inspirational to how we assess the environmental aspects in our method, there is not a single framework on which we built.
Finally, audit cards are suggested as a way to document all relevant aspects of an audit, including information about the auditors, target of the audit, the audit procedure, review and feedback possibilities for the audit, access and resources of the auditors, and integrity of the audit. Such audit cards might be a relevant extension of our work \cite{staufer_audit_2025}.

Another important strand of literature are international standards, such as the \textit{ISO/IEC standard 42005 on AI System Impact Assessments} \cite{ISO_IEC_42005}, \textit{ISO 26000} on guidance on \textit{Social Responsibility} \cite{ISO_260000} or the \textit{Global Reporting Initiative} (GRI) standards on \textit{Sustainability Reporting} \cite{GRI_standard}. Similarly relevant are guidelines by international institutions on the responsible and ethical use of AI, as for instance the \textit{Principles for Trustworthy AI} by the OECD \cite{OECD_Guidelines_on_Trustworthy_AI}, the \textit{Ethical Impact Assessment} (EIA) by the UNESCO \cite{UNESCO_Ethical_Impact_Assessment} or the EU \textit{Assessment List for Trustworthy AI} (ALTAI) \cite{EU_Assessment_List_for_Trustworthy_Artificial_Intelligence}.
Exemplary for international standards, the ISO/IEC standard 42005~\cite{ISO_IEC_42005} provides guidance on which topics should be included in an impact assessment. However, it does not formulate concrete questions for such an assessment, . 
Such topics include, among others, the purpose of the AI system, intended as well as unintended use and misuse of the system, and impacts on affected parties. They also cover technical information on the AI system, data, or algorithm. These topics are all important for the Impact-AI-method.
As a first example of international guidelines, the United Nations Educational, Scientific and Cultural Organization (UNESCO) formulated an Ethical Impact Assessment (EIA) \cite{UNESCO_Ethical_Impact_Assessment} for AI systems.
The EIA implements the UNESCO principles for ethical AI, which are safety and security, fairness, non-discrimination, and diversity, sustainability, privacy and data protection, human oversight and determination, transparency, explainability, accountability, and responsibility, and awareness and literacy. The document is relevant for the Impact-AI-method, since the UNESCO principles cover a wide range of topics which strongly overlap with public interest and sustainability concerns.
The EU Assessment List for Trustworthy AI~\cite{EU_Assessment_List_for_Trustworthy_Artificial_Intelligence}, developed by the High-Level Expert Group on Artificial Intelligence (AI HLEG) which was established by the European Commission, is another example of an important international guideline on AI systems. It outlines requirements and assessment questions for ethical and robust development and use of AI systems. 
It thereby covers similar topics as the EIA by the UNESCO, thus it is equally relevant to the Impact-AI-method.

\section{Development of the Impact-AI-Method}
\label{sec:method_development}

The Impact-AI-method aims to make the impact of AI systems for sustainability and public interest transparent, assessable, and measurable. This goal needs an interdisciplinary approach, because sustainability and public interest cannot be understood from a single perspective but only with different scientific expertise. Furthermore, it requires a qualitative approach to the audit that is based on social science methodology and a technical understanding of the AI systems to investigate. 
Furthermore, if the results of such an audit approach should have an impact beyond the academic community, the method should not only be scientifically sound but also be practically applicable and scalable.
Therefore, the method was developed in a transdisciplinary research project. The core research team has interdisciplinary expertise covering areas such as media studies, political theory, expertise in social science methodologies, knowledge on sustainability, and expertise in computer science and machine learning.
This is combined with the more practical knowledge provided by our project partners at \textit{Greenpeace} and \textit{Economy for the Common Good (ECOnGOOD)},\footnote{\url{https://www.greenpeace.de/} and \url{https://germany.econgood.org/}} two NGOs that are strongly concerned with sustainability.
Additionally, ECOnGOOD developed and uses its own audit procedure to understand the impact of private companies on the public interest, building on a self-developed matrix~\cite{harrach_transformation_2023}, therefore adding relevant practical experience to the research project.

To further validate and improve our method, and to include additional perspectives that are not covered by our project team, we set up a research council. It consists of eleven members from academia, civil society, and industry and covers different expertise, such as sustainability and AI, human rights, technology and law, and transformation~research.

The development process of the Impact-AI-method can be broadly summarized in the following six steps:

\begin{itemize}
    \item [1.] \textbf{Literature search:} We collected and reviewed relevant literature, focusing on auditing approaches and documentation of AI systems, relevant international AI frameworks and standards, as well as pertinent literature on sustainability, public interest and AI.
    An overview of the core literature is given in Tab.~\ref{tab:overview_literature} in the Appendix.
    \item [2.] \textbf{Extraction of indicators:} We extracted a first set of potential indicators of the Impact-AI-method from the previously reviewed literature.
    \item [3.] \textbf{Structure and fine-tuning:} We arranged the indicators into separate blocks that can be conducted as individual interviews and that have no (substantial) overlaps. We checked this version for critical gaps, redundancies, and understandability issues. In addition, this step was important to reduce the number of interview questions.
    \item [4.] \textbf{Assessment indicators:} We developed the catalog of assessment criteria and associated questions.
    \item [5.] \textbf{Mapping on life-cycle:} We aligned the assessment criteria with the AI life-cycle to create a matrix overview that allows to better understand \textit{when} which assessment indicator is most relevant.
    For simplicity, we used a life-cycle model with only four steps: \textit{planning and design}, \textit{data}, \textit{model}, and \textit{deployment and use}.\footnote{This life-cycle model is inspired by the SustAIn project and its online information material \url{https://sustain.algorithmwatch.org/en/step-by-step-towards-sustainable-ai/} as well as the connected publication \cite{rohde_broadening_2024}.}
    \item [6.] \textbf{Internal testing:} We conducted a test-audit with an AI project based at our institution. We then improved the method on the tripping points we discovered, e.g., in the interview flow and understandability of questions.
\end{itemize}

We took about 8 months to assemble the Impact-AI-method, including numerous update and feedback sessions in the core research team of interdisciplinary researchers, approximately 6 sessions with the partners from Greenpeace and ECOnGOOD, and one online session with the research council.
Additional small improvements to the method are expected through the audits of AI projects that we will conduct as part of our transdisciplinary research project.

\section{The Impact-AI-Method}
\label{sec:impact_ai_method}

\begin{figure}
    \centering
    \includegraphics[width=0.8\linewidth]{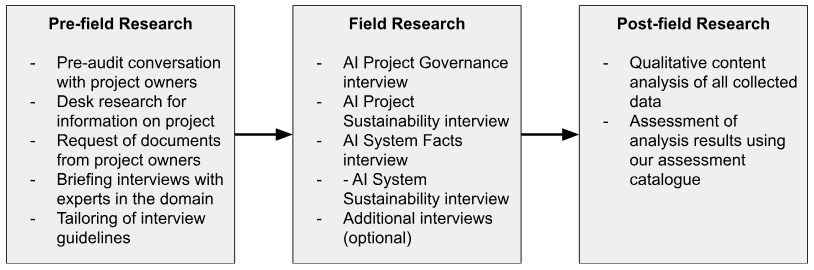}
    \caption{Overview of the research phases of the Impact-AI-method and main tasks.}
    \label{fig:audit_process}
\end{figure}

The following section describes the audit process of the Impact-AI-method in more detail, including an in-depth description of the four main audit interviews. The audit process can be divided into three phases: pre-field research, field research, and post-field research phase.
A graphical overview of the research phases indicating the main tasks in each phase is depicted in Fig.~\ref{fig:audit_process}.

\subsection{Pre-field Research Phase}

Having selected a case and having the agreement of the project owners to participate in the audit of their AI project, an exchange between project owners and auditors prior to the audit interview is important (`pre-audit conversation'). Firstly, a preliminary understanding of the workings of a project can be gathered, and secondly, the steps of the audit can be organized and the respective interview partners from the project team determined.
We also conduct desk research on publicly available information on the AI project. Common sources are the project website, project reports, research papers, news articles, and social media channels. These sources will be included in our data analysis in the post-field research phase. Since not all information is available through publicly accessible sources, we also directly request a list of documents from project owners. This includes detailed funding information, information on existing audits, and project reports. We also request the technical documentation of the AI models and datasets. 
Additionally, prior to field research, we arrange briefing interviews with independent experts in the respective application domain, providing us with relevant context knowledge that helps us to better contextualize the chosen AI project. 
Based on this information, as well as those provided by project owners and from our desk research, we then tailor the interview guide for audit interviews to the AI project. Certain questions may be unsuitable for a specific case, while others may be particularly important or even require further exploration. Additionally, new project-specific questions are added, e.g., based on the AI model type or a specific discourse topic in the application domain.

\subsection{Field Research Phase}

\begin{table}[htb]
    \begin{subtable}[t]{.5\textwidth}
    \scriptsize
    \centering
    \fbox{
    \begin{tabular}[t]{p{0.5cm}p{5cm}}
        \multirow{18}{*}{\rotatebox[origin=c]{90}{\texttt{AI Project Governance}}}
        & \textbf{Project information} \\
        & \quad 1. Organization and project profile \\
        & \quad 2. Roles and responsibilities \\
        & \textbf{Organizational goals and outcomes of the {AI} project} \\
        & \quad 3. Goal, visions and values \\
        & \quad 4. {Outcome} \\
        & \quad 4.1 Public interest \\
        & \quad 4.2 Sustainability \\
        & \quad 4.3 {Reflections on public interest and sustainability} \\
        & \quad 5. Outputs \\
        & \quad 6. Assessments \\
        & \quad 7. Strategies and obstacles \\
        & \quad 8. Trade-off regarding goals {or} impact \\
        & \quad 9. {Connection to the domain} \\
        & \textbf{Role of and decisions behind the AI system} \\
        & \quad 10. Proportionality of AI usage and role of AI \\
        & \quad 11. Decision process behind the AI system \\
        & \quad12. {Reflection on scarcity}  \\
        \midrule
        & \textbf{Social sustainability} \\
        & \quad 1. AI literacy \\
        & \quad 2. Self-determined use of AI system \\
        & \quad 3. Accessibility \\
        & \quad 4. Transparency \\
        & \quad 5. {Intervention and participation} \\
        & \quad 6. Stakeholders and co-design \\
        & \quad 7. Diversity \\
        & \quad 8. Potential harms \\
        & \quad 9. Liabilities and risk management \\
        & \quad 10. Protection mechanisms \\
        & \quad 11. {Working conditions} \\
        & \textbf{Environmental sustainability} \\
        & \quad 12. {Reflection on environmental impact} \\
        & \textbf{Economic sustainability} \\
        & \quad 13. Investments {for AI project} \\
        & \quad 14. Re-use and sharing \\
        & \quad 15. Compatibility and adaptability \\
        & \quad 16. Future perspective {of AI project} \\
        \multirow{-21}{*}{\rotatebox[origin=c]{90}{\texttt{\textcolor{gray}{AI Project Sustainability}}}}
        & \quad 17. Collaborations \\
    \end{tabular}
    }
    \end{subtable}%
    \begin{subtable}[t]{.5\textwidth}
        \centering
        \scriptsize
        \fbox{
        \begin{tabular}[t]{p{0.5cm}p{5cm}}
        \multirow{29}{*}{\rotatebox[origin=c]{90}{\texttt{AI System Facts}}} &
        \textbf{AI system {overview}} \\
        & \quad {1. Overview and development} \\
        & \textbf{{Model} facts} \\
        & \quad 2. Model date and version \\
        & \quad 3. Model details\\
        & \textbf{{Model} performance} \\
        & \quad 4. Relevant factors \\
        & \quad 5. Evaluation factors \\
        & \quad 6. Model performance measures \\
        & \quad 7. Approaches to uncertainty and variability \\
        & \quad 8. Embedding of AI model \\
        & {\textbf{Dataset}} \\
        & \quad {9. Description} \\
        &\textbf{Dataset collection} \\
        & \quad 10. Logistics \\
        &\textbf{Dataset composition} \\
        & \quad 11. Instances \\
        & \quad 12. Splits \\
        & \quad 13. Representation \\
        &\textbf{Dataset processing} \\
        & \quad {14. Preprocessing} \\
        & \textbf{Dataset maintenance} \\
        & \quad {15. Updates} \\
        & \quad {16. Storage} \\
        & \textbf{Ethical aspects} \\
        & \quad {17. Review processes and impact assessment} \\
        & \quad {18. Notification and consent duties} \\
        & \quad {19. Sensitive information} \\
        & \textbf{AI system reflection} \\
        & \quad {20. Reflection on alternatives} \\
        \midrule
        &  \textbf{Social sustainability} \\
        & \quad 1. Transparency \\
        & \quad 2. Non-discrimination \\
        & \quad 3. Privacy \\
        & \quad 4. Safety and security \\
        & \quad 5. Additional harms \\
        & \textbf{Environmental sustainability} \\
        & \quad 6. {Hardware overview} \\
        & \quad 7. Efficiency of the AI system \\
        & \quad 8. Energy and emissions \\
        & \quad 9. Materials \\
        & \quad 10. Certificates \\
        & \textbf{Economic sustainability} \\
        & \quad 11. Model Sharing \\
        & \quad 11.1 Sharing \\
        & \quad 11.2 Use \\
        & \quad 11.3 Compatibility \\
        & \quad 12. Dataset sharing \\
        & \quad 12.1 Sharing \\
        & \quad 12.2 Use \\
        \multirow{-22}{*}{\rotatebox[origin=c]{90}{\texttt{\textcolor{gray}{AI System Sustainability}}}} 
        & \quad 12.3 Compatibility \\
        \end{tabular}
        }
    \end{subtable}
    \caption{Overview of topics (\textbf{in bold}) and indicators (normal typeface) of core interviews of the Impact-AI-method. Left: overview of AI Project Governance and AI Project Sustainability Interview; 
    right: overview of AI System Facts and AI System Sustainability Interview. 
    For the interview guidelines that contain all questions, see the Appendix, Tab.~\ref{tab:interview_1_questions}, \ref{tab:interview_2_questions_part_1} and \ref{tab:interview_2_questions_part_2}; Tab.~\ref{tab:interview_3_questions_part_1}, \ref{tab:interview_3_questions_part_2}, \ref{tab:interview_4_questions_part_1} and \ref{tab:interview_4_questions_part_2}.}
    \label{tab:overview_topics_indicators}
\end{table}

In the field research phase, the research team conducts at least four interviews which form the core of the Impact-AI-method:
1) the \textbf{AI Project Governance Interview};
2) the \textbf{AI Project Sustainability Interview};
3) the \textbf{AI System Interview}; 
and 4) the \textbf{AI System Sustainability Interview}. 

All interviews cover a total of 68 topics and 218 questions. The first and second interviews are conducted with an organizational lead of the audited project team, and the third and fourth interviews are conducted with a technical lead. Although we acknowledge that a strict separation between technical and rather non-technical organizational interviews is not always possible, this separation has a strong practical advantage since we can do the interviews separately with different AI project team members and include different expertise.

All audit interviews follow are semi-structured interview approach \cite{tisdell_qualitative_2025}. Each interview guide is structured around a specific set of topics that summarizes several indicators (see Tab.~\ref{tab:overview_topics_indicators}). Indicators itself come with one or more questions.
The interview guides cover and organize the relevant topics and indicators extracted from the literature. At the same time, the semi-structured approach allows for openness and flexibility to respond to the interviewees' answers and to address additional topics that emerge during the interview.

If necessary and relevant for the specific case, the audit interviews are supplemented by \textbf{additional interviews} with users of or those affected by the AI system. Before each interview, the respective interviewees receive the data protection information that they need to agree to before the interview is conducted. All interviews are audio-recorded and transcribed.

\subsubsection{AI Project Governance Interview}
The goal of the interview is to gain a deeper understanding of the AI project's governance, referring, for instance, to processes, roles, or decision-making structures.
This includes gathering general \textbf{project information} with questions on the organizational structure of the organization behind the AI project and on how the project is embedded in it (indicators 1-2). 
For the topic of \textbf{organizational goals and outcomes of the AI project}, we ask questions on the project's self-defined purpose and its theory of change\footnote{A theory of change is a method that explains how a given intervention, or set of interventions, are expected to lead to a specific development change, drawing on a causal analysis based on available evidence \cite{UNDG}.} (indicators 3-5). Understanding such a theory of change thereby contributes to assessing the project's goals, intended outcomes and concrete outputs. Further questions are directed at providing insights on project management and governance.
We use reflective questions to assess the alignment of the project's purpose with public interest and sustainability (e.g. \enquote{Do you see your project serving an interest that could be understood as a public interest? If yes, how?}) (indicators 4.1-4.3).
The following topic \textbf{role of and decisions behind the AI system} provides insights in the decision-making processes behind and the development of the AI project. In this part, we ask questions to comprehend the rationale for using an AI system to reach the project's purpose (indicator 10). We also directly ask for decisions with respect to the AI system (\enquote{What were relevant decisions about the architecture of your system (e.g., the languages in which the system works, the design of the interface?}) and the processes behind these (\enquote{Based on which arguments did you make these decisions and what was the decision-making process (e.g., stakeholder involvement)?}) (indicator 11). 

\subsubsection{AI Project Sustainability Interview}
The objective of this interview is to understand the representation of the three sustainability dimensions within the AI project.
The first part of this interview covers different indicators on the topic of \textbf{social sustainability} of the AI project. One of these indicators is accessibility, which collects questions on accessibility of the AI application, including considerations related to user groups with different or special needs (indicator 3). The questions are linked to the project's implication for equality, a foundational element of social sustainability. In addition, the interview consists of questions about the transparency of the overall project and the AI system in use (indicator 4). Questions on possibilities of participation are included, addressing possible communication channels for feedback or complaints about the AI project by its users or the general public (indicator 5). The questions on co-design cover whether relevant stakeholder groups were involved in the development of the AI project and system (indicator 6). These criteria--transparency, participation, and stakeholder involvement--are integral for understanding the implementation of processes supporting an orientation towards the public interest. Moreover, the interview includes reflections on potential harms of the AI project and possible protection measures taken (indicators 8-10). The answers to this question provides insights on whether the project owners critically assessed and prepared for negative implications before using the AI system. 
The topic of \textbf{environmental sustainability} of the AI project is addressed by asking reflective questions about general considerations regarding the environmental impacts of AI systems. This considers AI systems in general, but also the specific AI project under investigation (indicator 12).
The last topic of the interview, \textbf{economic sustainability} of the AI project, covers the necessary financial expenses and project funding (indicator 13). Also, we ask questions on the implementation of the system in other contexts including feasibility of system adoption, compatibility, and adaptability (indicators 14-15). Finally, we include questions about plans for sustaining the project in the future, including collaborations with other organizations or initiatives (indicators 16-17).

\subsubsection{AI System Facts Interview}

The main goal of this interview is to understand the functionality of the AI system.
Such a system consists of at least one AI model, as well as at least one dataset.
The interview starts with a set of general questions on the \textbf{AI system} and its development, to create an overview for the interviewer.
This is followed by several topics on the \textbf{AI model} (indicators 2-8), which are strongly based on~\cite{DBLP:conf/fat/MitchellWZBVHSR19}.
First, we ask about general information about the model (indicators 2-3).
Model performance (indicators 4-8) covers factors, i.e., the factors by which performance of the model may vary (indicators 4-5). For example, the performance of an NLP model often varies by language \cite{DBLP:journals/corr/abs-2508-17162}. 
We also ask for the specific measures that the organization uses to evaluate the AI model and information about how often such an evaluation is repeated (indicator 6). We cover approaches towards uncertainty and variability (indicator 7), and the embedding (indicator 8). Embedding thereby refers to the software embedding of the AI system, i.e, a web environment. We collect its technical specifications, as well as information about data that may be collected through it.
Regarding the topics related to the \textbf{dataset} (indicators 9-16), we build on \cite{,DBLP:journals/cacm/GebruMVVWDC21}, and collect information about the collection process (\enquote{logistics}), for example, how the data was collected, and over which time frame.
We collect information about its composition (indicators 11-13), for example, what each instance represents and how many instances the dataset has. Also, we collect details about the dataset splits and the representation of the dataset regarding the task.
Additional indicators (indicators 14-16) concern the dataset preprocessing and its maintenance, including how often it is updated and where it is stored.
Last, we also include a set of questions about the ethical aspects of the dataset, concerning review processes and impact assessments conducted (indicator 17), notification and consent duties (indicator 18) and sensitive information (indicator 19).
We close with a reflection question on alternatives to the \textbf{AI system} developed (\enquote{Is there anything that you would do differently about the AI system, if you developed it~again?}).

\subsubsection{AI System Sustainability Interview}

The goal of this interview is to comprehensively understand the different sustainability dimensions of the AI system, with a focus on their technical aspects.
Starting with the topic of \textbf{social sustainability} of the AI system (indicators 1-6), we cover transparency, non-discrimination, privacy, as well as safety and security aspects, which are all part of several high-level ethics guidelines on AI systems \cite{EU_Assessment_List_for_Trustworthy_Artificial_Intelligence,UNESCO_Ethical_Impact_Assessment,OECD_Guidelines_on_Trustworthy_AI} and that are critical for their proper functioning without creating social harms.
We also ask about \enquote{additional harms} (indicator 6), to cover those that do not occur in every AI model or application, e.g., hallucination harms of LLMs.
The topic of \textbf{environmental sustainability} with respect to the AI system (indicators 7-11) starts with a question about the hardware in the AI project, to gain an overview of the infrastructure behind the project.
We also cover the efficiency of the AI system (including, complexity and model size), which is critical for resource consumption. 
In this topic, we ask both reflection questions and specific questions about measuring energy usage and emissions, as well as about the material use behind the infrastructure. This is concluded by questions about plans after decommissioning the AI system and hardware certificates.
The \textbf{economic sustainability} of the AI system as a topic (indicators 12-13) includes the sharing of models and datasets, covering aspects such as the details of who to contact about the shared model and dataset and with which other models, datasets, or programming environments they are compatible with.
The question of usage is also very important--we do not only ask about the possible further use of model and dataset, but also about their potential out-of-scope usage.
Public availability of an AI system for re-use is an important sustainability aspect, as it reduces overall resource consumption and the requirements for other actors to utilize the AI system \cite{rohde_broadening_2024}.

\subsubsection{Additional Interviews}

Depending on the audited AI project, additional semi-structured interviews are conducted with organizations or individuals using the AI system, or with those who are affected by the system. To this end, purposive sampling is deployed for interview selection \cite{etikan_comparison_2016}.
The interview guide consists of selected excerpts from the four core audit interviews focused on the application of the system and respective impacts. Additional questions may be added depending on stakeholders and context. These additional interviews, depending on the specific use case scenario, support a better understanding of the impact of the AI system in concrete application contexts. 

\subsection{Post-field Research Phase}

In the post-field research phase, the collected data, including interview transcripts and documents, is analyzed using the software MaxQDA and following a qualitative content analysis. 
The assessment criteria (see Tab.~\ref{tab:assessment_indicators} in the Appendix) form the code framework, i.e., the categories used to label the interviews in order to systematically analyze the data.
Using an extractive and structured approach \cite{Glaser_Laudel_2010, Kuckartz_2018}, we can supplement these deductive codes by inductive codes emerging in the course of the analysis. 
The combination of such a systematic examination and openness to emerging codes ensure a comprehensive understanding of the impact of AI projects on sustainability and public interest at various~levels.

After data analysis, the results will be assessed. For the assessment of the AI projects, we developed 32 assessment criteria, each of them associated with a question.
Each assessment question is also paired with one or more questions in the audit interview questionnaire, to make clear which statements (and thus audit results) the specific assessment question depends on. 
Furthermore, each criterion is sorted into one of the six following clusters: governance, transparency and participation, model and data facts, as well as social, environmental and economic dimensions of sustainability. 
Once the answers to the assessment questions are collected, we judge these answers on a scale of five. 
Here, we use an extended traffic light model in which the colors indicate the performance of the AI project, ranging from green (for positive) to red (for negative). The objective of such an assessment is twofold. First, it provides a learning experience for project owners, highlighting aspects within the organization and the AI system where improvements can be made or where performance is already strong. Second, the results of the assessment facilitate knowledge transfer to civil society and other sectors such as public administration, as they are presented in a clear and easily accessible way.

\section{Discussion}
\label{sec:discussion}

The following three points require further scholarly discussion: 1) the relationship between our Impact-AI-method and compliance focused audits; 2) comparability of audit results for very different use cases; and 3) the impact our and other audit methods can have.

The first point considers the \textbf{relationship between our approach and compliance-focused audits}.
We see our audit as a complementary approach: legally focused audits aim at assuring compliance and achieving transparency and accountability in regards to potential harms that also constitute violations of applicable laws, such as privacy and data protection, anti-discrimination requirements and risk management obligations  \cite{lam_framework_2024, lacmanovic_skare_2025}. While acknowledging the importance of this assurance of legal compliance, the Impact-AI-method aims at producing insights for the societal debate of AI's role for a just and sustainable development of society. This goes, at least in some dimensions, beyond what is required by the law, especially since our method focuses around positive societal objectives instead of risks and harms to society.
This difference also shows in the audit design, as our method operates mainly with open interview questions.
However, we do see legal audits and regulatory compliance as strictly relevant, as they represent the status quo of negotiated minimum standards that each AI project needs to live up to.

Another issue is how to tailor our method to different use cases and still ensure the \textbf{comparability of audit results}. We already used the Impact-AI-method for two audits (at the time of the submission) in which we applied the method to concrete use cases. A core learning from this application is that even if AI projects claim to have similar objectives (e.g., using AI to support democratic processes), tailoring to the specific use case is necessary for meaningful results.
Most of the questions of the Impact-AI-method are the same for all projects, while a small amount need to be adapted for each case. For example, these questions address the harms or unintended side effects that arise from the AI project
or the technical details specific to the AI system.
Despite these tailored questions, the audits are comparable: most of the questions remain the same, and the assessment criteria (and thus topics) are also not changed.
We chose such a tailoring approach over others, as we believe it produces more insightful data than a less flexible, general approach but still allows to produce meaningful results for a debate of the criteria.

Last, we would like to discuss \textbf{the impact of our and other AI audits}. As Birhane et al. \cite{DBLP:conf/satml/BirhaneSOVR24} point out \enquote{[s]ubsequently, more often than not, audits are seen as an academic, intellectual exercise rather than practices directly linked to real world consequences.} 
We intend to broaden the scope of the Impact-AI-method beyond the academic realm. We believe that with the help of our method, we can derive demands for concrete individual change within AI projects. At the same time, by creating a larger set of comparable audit cases as part of future work, we see a possibility to arrive at more systematic demands for change. Part of this are also the assessment criteria developed to rate the performance of the AI projects and the facilitation of knowledge transfer outside of the academic realm. 
As a strategy to address the challenge of creating concrete impact on a practice level, we developed communication tailored to non-academic audiences, e.g. in the form of the matrix (see Fig.~\ref{fig:placeholder} in the Appendix). In addition, we developed a theory of change for the impact of the research project to ensure a structured evaluation of our own intended outcomes. This is a strategic document to spell out the implications, the intended goal, achievable outcomes, and concrete outputs for our research project, which goes beyond the Impact-AI-method. 
Our method is intended to be applied repeatedly in the future, beyond our research project, as an independent audit. Potentially, this can be done by other researchers and even non-academic actors. To this end, in the course of our research project, we will develop an organizational structure to support the longer-term use of our method as well as a structure to ensure its qualitative stability (or even improvement), including, for instance, train-the-trainer modules to transfer method knowledge to other stakeholders. 

\section{Conclusion and Outlook}
\label{sec:conclusion_outlook}

AI systems are increasingly used with the intention of driving just and sustainable development, often labeled AI `for good'. However, there is a lack of transparency in the goals behind such AI projects, as well as insufficient evaluation of their actual impact.
To this end, this paper provides two contributions: by combining and operationalizing the theoretical concept of public interest and sustainability, we first provide a framework to theoretically assess AI systems in regards to the objections and tensions for a just and sustainable development.
Building on this understanding, we present the Impact-AI-method, an audit method that can serve as a tool to gain evidence of the actual impact of an AI system.
Part of this method is an assessment catalog to rate the results of the Impact-AI-method. This translation of the audit results into an easily understandable overview makes the results accessible to an audience beyond academia.

As part of the research project, we will apply the Impact-AI-method to 15 selected AI projects and thereby not only gain evidence from the field but also iteratively improve the method based on the findings of its application.
In addition, we plan to supplement the method with experiments on models and data. Whether and to what extent they can be conducted depends both on the access to the AI models and datasets, as well as the type of AI project audited. 
Other future steps of improvement include, for instance, developing scenarios to further improve and support the assessment criteria, for a harmonized assessment practice. One way to do this is by describing the worst- and best-case scenarios for each assessment option.

The overall goal of our research is to provide more evidence on the impact of AI projects for sustainability and public interest and therefore also to the accountability around the use of AI.
By this, we want to contribute to the discussion on the societal change AI systems support–-a change that is not inevitable but builds on human implementation, purposeful design, and decision-making.

\newpage
\section*{Endmatters}


\subsection*{Generative AI Usage Statement}
When preparing the paper, we did not use generative AI tools to generate any of the text.
However, we used the Writeful Overleaf AI Editing Assistant to support grammar editing and proof-reading of the manuscript.


\printbibliography


\newpage
\appendix

\begin{table}[h]
    \centering
    \scriptsize
    \begin{tabular}{cp{11cm}p{2cm}}
    \toprule
    Source & Title & Type \\
    \midrule
    \cite{DBLP:conf/fat/MitchellWZBVHSR19} & Model cards for model reporting & peer-reviewed \\
    \cite{DBLP:journals/cacm/GebruMVVWDC21} & Datasheets for datasets & peer-reviewed \\
    \cite{rohde_broadening_2024} & Broadening the perspective for sustainable artificial intelligence: sustainability criteria and indicators for Artificial Intelligence systems & peer-reviewed \\
    \cite{solaiman2024evaluatingsocialimpactgenerative} & Evaluating the Social Impact of Generative AI Systems in Systems and Society & peer-reviewed \\
    \cite{zicari_how_2022} & How to assess trustworthy AI in practice & report \\
    \cite{zuger_ai_2023} & AI for the public. How public interest theory shifts the discourse on AI & peer-reviewed \\
    \cite{zuger_civic_2022} & Civic coding. Fundamentals and empirical insights for supporting AI that serves the public interest & report\\ 
    \cite{rohde_nachhaltigkeitskriterien_2021} & Nachhaltigkeitskriterien für künstliche Intelligenz~(*) & report \\
    \cite{UNESCO_Ethical_Impact_Assessment} & UNESCO Ethical impact assessment: a tool of the Recommendation on the Ethics of Artificial Intelligence & high-level guideline \\
    \cite{OECD_Guidelines_on_Trustworthy_AI} & OECD Guidelines on Trustworthy AI & high-level guideline \\
    \cite{EU_Assessment_List_for_Trustworthy_Artificial_Intelligence} & EU Assessment List for Trustworthy Artificial Intelligence (ALTAI) for self-assessment & high-level guideline \\
    \cite{ISO_IEC_42005}  & ISO/IEC 42005 & standard \\
    \cite{ISO_260000} & ISO 26000 & standard \\
    \cite{GRI_standard} & Global Reporting Initiative (GRI) & standard \\
    \cite{harrach_transformation_2023} & Transformation von Unternehmen mit der Gemeinwohl-Ökonomie: Wissen, Werkzeuge und Motivationen zur nachhaltigen
    Organisationsentwicklung (*) & book \\
    \bottomrule
    \end{tabular}
    \caption{Overview of core literature behind the Impact-AI-method. 
    (*) available only in German.}
    \label{tab:overview_literature}
\end{table}

\begin{table}[h]
    \scriptsize
    \centering
    \begin{tabular}{p{4cm}p{10cm}}
    \toprule
    Assessment Criteria & Assessment Question \\
    \midrule
    \rowcolor{teal!10}
    {\textbf{Governance}} & \\
    1. Goal and public interest & Does the described goal of the AI project create a comprehensible connection to conceptions of public interest? \\
    2. Goal and sustainability & Does the described goal of the AI project create a comprehensible connection to at least one dimension of sustainability? \\
    3. Impact achievements & Is there convincing evidence supporting the claim that the AI project has achieved envisioned outcomes? \\
    4. Theory of change & Is a theory of change convincing and well-reflected? \\
    5. Proportionality of AI usage and role of AI & Is the use of AI proportional and helpful to the project's goal and is the use of AI reflected (also considering alternatives)? \\
    6. Decision processes AI system & Does the AI project convincingly describe a process of decision-making about the project's purpose and design that represents relevant stakeholders and a comprehensive connection to public interest? \\
    7. Liability, risk assessment and management & Are there appropriate responsibilities and response mechanisms in cases of liability? \\
    \rowcolor{teal!10}
    \textbf{Transparency and participation} & \\
    8. Transparency & Does the AI project sufficiently create meaningful transparency about its goals, systems (incl. AI use), strategies and intended outcomes? \\
    9. Open for validation & Does the project convincingly allow for third party validation (through audits or otherwise)? \\
    10. Intervention / Participation & Is there convincing interest in third party critique or opinions and sufficient channels to allow entry points for participation? \\
    11. Stakeholders and co-Design & Are there appropriate procedures in place to include relevant stakeholders in the design, deployment and use of the AI system? \\
    \rowcolor{teal!10} 
    {\textbf{Model and dataset facts}} & \\
    12. Performance & Is the AI model sufficiently evaluated in detail? \\
    12.1 Factors & Are the relevant factors that influence the performance of the AI model (incl. bias) evaluated? \\
    12.2 Measures & Are performance measures appropriate to the AI model type and task? \\
    12.3 Adequacy to project purpose & Is the performance of the AI model adequate to achieve the envisioned purpose of the system? \\
    12.4 Uncertainty and Variability & Is the uncertainty and variability in the performance of the AI model sufficiently taken into account? \\
    13. Data composition and preprocessing & Is the dataset well suited for the task (representation, size, appropriate data preprocessing)? \\
    14. Data maintenance & Is there a suitable plan for maintaining the dataset? \\
    15. Ethical aspects of data & Are ethical aspects of the dataset sufficiently taken into account? \\
    \rowcolor{teal!10}
    {\textbf{{Social sustainability}}} & \\
    \rowcolor{pink!10}
    \textbf{Model and Data} & \\
    16. Transparency & Does the AI project create meaningful and sufficient transparency about its AI and data use (on a technical level)? \\
    16.1 Documentation & Is the AI model and data sufficiently well documented? \\
    16.2 Model functionality & Is the functionality of the AI model sufficiently well explained (e.g., using an XAI method)? \\
    17. Non-discrimination & Was the potential of discrimination through the AI model assessed and steps to prevent such discrimination taken (if applicable)? \\
    18. Privacy & Does the AI project sufficiently protect privacy of individuals using the AI model (if applicable)? \\
    19. Safety and security & Is the AI system (model and data) sufficiently protected against different types of attacks? \\
    \rowcolor{pink!10}
    \textbf{Organization/Project} & \\
    20. Self-determined use & Do users make a self-determined choice to use the AI system? \\
    21. AI literacy & Does the AI project describe efforts or concrete measures to improve users' AI literacy? \\
    22. Accessibility & Does the AI project convincingly describe efforts or create measures to create accessibility? \\
    23. Diversity & Are diverse perspectives adequately reflected and supported in the AI project? \\
    24. Harmreflection and harmmanagement & Does the AI project convincingly reflect potential harm and take appropriate measures to prevent it? \\
    25. Manifestation of good working conditions & Is the AI project convincingly committed to establishing good working conditions within their own organization as well as for workers along the AI life-cycle? \\
    \rowcolor{teal!10}
    {\textbf{Environmental sustainability}} & \\
    26. Efficiency of AI system & Is the efficiency of the AI system and its impact on the environment sufficiently considered? \\
    27. Energy and emissions & Is energy use and emissions of the AI system and its impact on the environment sufficiently considered? \\
    28. Materials & Is material use of the AI system and its impact on the environment sufficiently considered? \\
    29. Certificates & Are certificates for hardware used or considered for use? \\
    \rowcolor{teal!10}
    \textbf{ECONOMIC SUSTAINABILITY} & \\
    \rowcolor{pink!10}
    \textbf{Model and Data} & \\
    30. Model sharing: general & If suitable in the context -- is the AI model made available for further use or are considerations about model sharing being made? \\
    30.1 Model sharing: use & If shared, is consideration taken of possible adverse use? \\
    30.2 Model sharing: compatibility & If shared, is consideration taken w.r.t. compatibility? \\
    31.  Dataset sharing: general & If suitable in the context -- is the data shared for further use or are considerations about dataset sharing being made? \\
    31.1 Dataset sharing: use & If shared, is consideration taken of possible adverse use? \\
    31.2 Dataset sharing: compatibility & If shared, is consideration taken w.r.t. compatibility? \\
    \rowcolor{pink!15}
    \textbf{Organization/Project} & \\
    32. Future perspective and collaboration & Are there efforts to establish collaborations and to sustain the AI project and the fulfillment of its objectives in the future? \\
    \bottomrule
    \end{tabular}
    \caption{Assessment criteria and assessment questions to analyze the {results} of the interviews.}
    \label{tab:assessment_indicators}
\end{table}

\begin{figure}
    \centering
    \includegraphics[angle=0, width=0.95\linewidth]{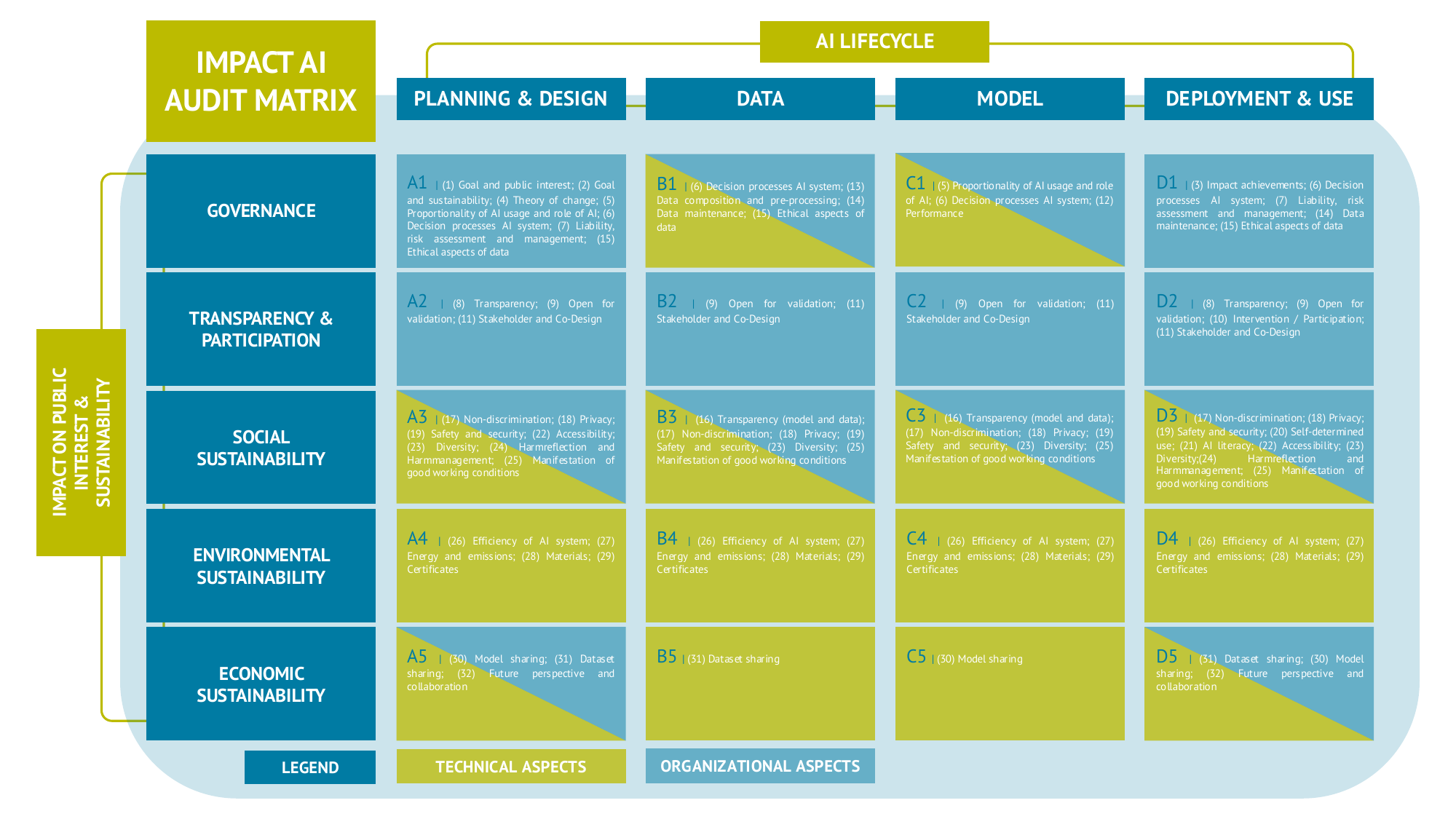}
    \caption{Visualization of the assessment criteria as matrix.}
    \label{fig:placeholder}
\end{figure}

\begin{table}[]
    \scriptsize
    \centering
    \begin{tabular}{p{3cm}p{6cm}p{5cm}}
        \toprule
        Topic/Indicator & Question & Follow-up/optional question \\
        \midrule
        \rowcolor{teal!10}
        \textbf{Project information} && \\
        1. Organization and project profile & - Since when does your AI project exist? & \\
        & - How many people work for the project? & \\
        & - What is your organizational structure and governance and how is this project embedded into it? & \\
        & - What jurisdiction applies for your project and what documentation is therefore necessary? & \\
        2. Roles and responsibility & - Who has responsibility for the major work streams within this project. 
        Include a full description of roles and responsibilities within the team.
        & \\
        && - Who has ultimate decision-making authority within the project team responsible for this AI system? \\
        && - Are there contact persons for ethical and social matters for the project? \\
        \midrule
        \rowcolor{teal!10}
        \multicolumn{3}{l}{\textbf{Organizational goals and outcomes of the AI project}} \\
        3. Goal, visions and values & - Which societal goal do you envision to achieve with your project? & \\
        & - What values are relevant drivers of your work? & \\
        4. Outcomes & - What are the concrete outcomes of this project? 
        If the aim is to address a specific problem, please specify the problem you are trying to solve. & \\
        & - How did the project's goals and its strategy to use AI evolve? & \\
        & - Why is this AI system the right solution to the specific problem you aim to solve? & \\
        4.1 Public interest & - Do you see your project serving an interest that could be understood as a public interest? If yes, how? & \\
        & - Can you explain your understanding of public interest? & \\
        4.2 Sustainability & - Does your project serve sustainability in one or more of the commonly addressed dimensions: social, economic and ecologic sustainability? If yes, how? & \\
        4.3 Reflections on public interest and sustainability & - Do you think that AI systems which are inherently built on practices of extraction and exploitative working conditions could ever be in the public interest or sustainable? & \\
        && - How do you reflect on your use of LLMs that are built on these practices? \\
        5. Outputs & - What are the specific outputs of your project, meaning deliverables or means required to achieve your outcomes and goals? & \\
        6. Assessment & - How effective has the project been in supporting to achieve the envisioned outcomes? & \\
        & - Which role did the AI system play for achieving these outcomes? & \\
        & - What do you think are relevant signs of the success of your work? & \\
        & - Did you observe any unintended - positive or negative- outcomes? & \\
        && - What did not work as intended? \\
        & - Are there any measures to monitor the outcomes and outputs of your work. If yes, which? & \\
        7. Strategies and obstacles & - What do you think are factors for your project to be successful in achieving your outcomes? & \\
        & - What do you see as obstacles to your project's success and impact? & \\
        & - What do you see as relevant strategies you realized to react to obstacles? & \\
        8. Trade-off in goals and impact & - Do you see the goals of your project aligned or in conflict with each other? & \\
        && - How do you decide on trade-offs in regard to these conflicting goals? 
        \\
        9. Connection to the domain & [Questions specific to case and application domain] \\
        \midrule
        \rowcolor{teal!10}
        \multicolumn{3}{l}{\textbf{Role of and decisions behind the AI system}} \\
        10. Proportionality of AI usage and role of AI & - Has careful consideration been given to non-algorithmic options which may be used to achieve the same outcomes? & \\
        && - If so, why is the option involving an AI system favored? \\
        & - Which proportion has AI technology compared to human or social factors in solving the problem you address? & \\
        & - How were affected people involved in the decision on using AI as a tool? & \\
        11. Decision process behind the AI system & - What were relevant decisions about the architecture of your system (e.g., the languages in which the system works, the design of the interface?) & \\
        & - Based on which arguments did you make these decisions and what was the decision-making process (e.g., stakeholder involvement)? & \\
        12. Reflection on scarcity & - Imagine your AI project under a condition of scarcity in regards to natural resources, money and technical infrastructure. What would you have done differently under this condition?& \\
        \bottomrule
    \end{tabular}
    \caption{Interview 1: topics, indicators, questions, and follow-up/optional questions.}
    \label{tab:interview_1_questions}
\end{table}

\begin{table}[]
    \scriptsize
    \centering
    \begin{tabular}{p{3cm}p{6cm}p{5cm}}
        \toprule
        Topic/Indicator & Question & Follow-up/optional question \\
        \midrule
        \rowcolor{teal!10}
        \textbf{Social sustainability} && \\
        1. AI literacy & - Which competencies or knowledge are necessary to use the system? & \\
        & - Have any means been put in place to help educate users and impacted groups about this system and the reason behind its deployment? & \\
        && - Can you provide an example or evidence that any measure taken had a positive effect on AI literacy of users? \\
        2. Self-determined use & - Are users self-motivated to use the system? & \\
        && - What degree of optionality will users have? \\
        3. Accessibility & - How did you consider aspects of accessibility for your application and did you have specific user groups in mind for these consideration? & \\
        && - Did you assess whether the AI system is usable by those with disabilities? \\
        && - Did you assess whether the AI system is usable by those with a precarious economic situation or low internet connection? \\
        && - Is your application explained in simple language and simple to understand? \\
        4. Transparency & - Which information about the project is available for which target groups and in which depth? & \\
        && - Which information about your AI system is publicly available? \\
        & - Is it clear from using your system that AI is used? \\
        5. Intervention and participation & - How can people engage with you, comment on your work or express their concern with your project? & \\
        && - How often does this happen and what was the feedback? \\
        && - Is there a procedure in place to investigate claims raised about the system by the general public, researchers or the media? If yes, please elaborate on the procedure(s). \\
        && - Can individuals impacted by the AI system submit complaints to the project? \\
        && - How can individuals make a request for an explanation of how a decision was made and how are they made aware that this is possible? \\
        && - What are the processes of such a request, who receives it and how do you respond? \\
        6. Stakeholders and co-design & - Who do you see as relevant stakeholders for your project? & \\
        & - Who are users of your system? & \\
        && - Are any of these stakeholders (or user groups) more vulnerable in any regard to impacts of the project? \\
        & - Which stakeholder groups are involved or consulted during the development, deployment and use of the AI system? & \\
        & - How much did stakeholder input influence project development and decision-making? Can you give an example of how your work was influenced by the perspective of affected stakeholders? & \\
        && - What objective do you have for engaging these stakeholders? \\
        && - How often do you plan to repeat processes of stakeholder involvement? \\
        && - What resources are available to realize stakeholder engagement and what time/financial constraints may limit involvement? \\
        && - Which modes of stakeholder engagement would have been more appropriate, given these constraints did not exist? \\
        7. Diversity & - Has consideration been given to the diversity of the AI project team, 
        and how diversity might effect biases? 
        & \\
        & - Are there any measures or practices in place to encourage diverse perspectives and people working in your organization? & \\
        \bottomrule
    \end{tabular}
    \caption{Interview 2, part 2: indicators, questions and follow-up/optional questions.}
    \label{tab:interview_2_questions_part_1}
\end{table}

\begin{table}[]
    \scriptsize
    \centering
    \begin{tabular}{p{3cm}p{6cm}p{5cm}}
        \toprule
        Topic/Indicator & Question & Follow-up/optional question \\
        \midrule
        8. Potential harms & - Which potential harm or negative side effects could result from your work? & \\
        && - Could the AI system and its application impact fundamental human rights? \\
        && - Is the system intended to be used for mass surveillance? Could the system be adapted for mass surveillance by other actors? If so, have measures been put in place to safeguard against this? \\
        && - Are the expected impacts irreversible or difficult to reverse? \\ 
        && - Could the expected impacts involve life and death decisions? \\
        && - Does your system contain any addiction-enhancing mechanisms? If yes, which? \\
        && - Could there be potential harm through over-reliance on the AI system, especially in high stakes decisions? \\
        && - Could any job loss result from your project? \\
        9. Liabilities and risk management & - Are there regulations and/or internal rules on liability aspects for your project? & \\
        && - In a liability case, who is the responsible person/entity in your project? \\
        & - Do you assess the liability risks of your system. If yes, how? & \\
        && - Did you consider risk [specific to case and application domain] for your project? \\
        10. Protection mechanisms & - What are the mechanisms in place if the AI system takes a wrong decision or creates concerning outputs [or specific harms mentioned before by the interviewee]? & \\
        && - Do your protection mechanisms allow for human oversight? \\
        && - Do you deploy any measures to ensure that your system does not create biased or discriminatory outcomes that are beyond technical measures? \\
        11. Manifestations of good working conditions & - How were working conditions along the AI life-cycle considered in the project? & \\
        & - Describe the working conditions within your project and assess their qualities and shortcomings.  \\
        & - Describe your working culture, your management culture and power relations in your project. & \\ 
        & - Do you have a code of conduct and if yes, can you forward it to us? & \\
        & - How was this code of conduct produced and how are people made aware of it? & \\
        & - Is there a procedure or person approachable for complaints or conflict resolution in your project or organization? & \\
        \midrule
        \rowcolor{teal!10}
        \textbf{Environmental sustainability} && \\
        12. Reflections on environmental impact & - How do you see the environmental impact of AI systems in general and how do you reflect on your own system's impact in this regard? & \\
        && - Did your reflection lead to any type of action? \\
        & - How do you see your responsibility in regards to the impacts of the AI industry on nature? & \\
        \midrule
        \rowcolor{teal!10}
        \textbf{Economic sustainability} && \\
        13. Investment for AI project & - What investment was necessary to realize your project? & \\
        & - What are the sources of your funding? & \\
        && - How much did you spent on: 1) storing data; 2) computing hardware; 3) hosting/inference? \\
        && - How much did you spent on other costs (e.g., wages or office costs)? \\
        14. Re-use and sharing & - Is your model and data already used in other contexts? & \\
        & - Have others expressed interest in your model and data? & \\
        & - Could your system be implemented by others easily and how feasible would an adoption be? & \\
        & - Did you engage in any effort to make the adoption of your system easier? & \\
        15. Compatibility and adaptability & - What steps would be necessary and what would be the work invest to adopt the technology to other use cases? & \\
        & - Would the application be able to scale and to which degree? & \\
        16. Future perspective of AI project & - What is your plan to keep the project available for the next 3-5 years? Please elaborate on concrete measures to maintain the project. & \\
        & - How do you want to ensure that the project stays true to its societal goals in future scenarios, even if ownership changes? & \\
        17. Collaboration & - Do you have active collaborations with other organizations or initiatives and if yes with which goal are these pursuits? & \\
        && - Are there other types of knowledge exchange that your pursue? \\
        \bottomrule
    \end{tabular}
    \caption{Interview 2, part 2: indicators, questions and follow-up/optional questions.}
    \label{tab:interview_2_questions_part_2}
\end{table}

\begin{table}[]
     \scriptsize
    \centering
    \begin{tabular}{p{3cm}p{6cm}p{5cm}}
        \toprule
        Topic/Indicator & Question & Follow-up/optional question \\
        \midrule
        \rowcolor{teal!10}
        \textbf{AI system {overview}} && \\
        1. {Overview and development} & - Describe the structure of your AI system. & \\
        & - Describe the main steps in the development of the AI system. & \\
        \midrule 
        \rowcolor{teal!10}
        \textbf{{Model} facts} && \\
        2. Model date and version & - When did you start implementing the model and which is the current version you use? & \\
        3. Model details & - Comment: Do you use one model, or do you combine several models? If the latter is the case, we need the following information for all models. & \\
        & - What type of a model is it? & \\
        && - What type of a model is it according to the following classification: 1) symbolic or sub-symbolic; 2) supervised, unsupervised or reinforcement; 3) generative or discriminative? \\
        & - What is the architecture of your model? & \\
        && - Generative/NLP: Which base model do you use? \\
        & - What is the task of the model? & \\
        & - How many parameters has it and can you describe the parameters and their effect on the model? & \\
        & - What type of data can it be applied to? & \\
        & - Which programming language and which libraries did you use? & \\
        & - Do you plan to retrain your model? If yes, how often? & \\
        \midrule
        \rowcolor{teal!10}
        \textbf{Model performance} && \\
        4. Relevant factors & - Which are the relevant factors on which you expect that the model performance varies?
        & \\
        5. Evaluation factors & - Which are the factors by which you report model performance? 
        & \\
        6. Model performance measures & - How did you evaluate the performance of your model? & \\
        && - Generative AI: Which benchmark did you use to evaluate your model? \\
        && - Discriminative AI: If you use a threshold, what is the value and how was it determined? \\
        & - Can you provide these evaluations? & \\
        && - How do the performance results relate to the different factors? \\
        & - Do you continuously monitor performance of your system? & \\
        && - How often do you repeat the performance evaluation? \\
        7. Approaches to uncertainty and variability & - How did you measure the uncertainty and variability of the performance of your model, and what are the values? & \\
        8. Embedding of AI model & - How is your AI model embedded in your app/web environment? & \\
        & - What are the technical aspects of the environment (e.g., the programming language or the server it is hosted on)? & \\
        & - Do you collect any additional data through this embedding? If yes, which and how do you protect it (e.g., passwords to access the app)? & \\
        \midrule
        \rowcolor{teal!10}
        \textbf{Dataset} && \\
        9. {Description} & - Describe the dataset shortly. & \\
        \midrule
        \rowcolor{teal!10}
        \textbf{Dataset collection} && \\
        10. Logistics & - Did you collect the data specifically for this task or did you have other tasks in mind? & \\
        & - How did you collect the data? & \\
        & - If the dataset is a sample from a larger set, what was the sampling strategy? & \\
        & - Who was involved in the data collection and was there a compensation? & \\
        & - Over what time frame was the data collected? & \\
        \bottomrule
    \end{tabular}
    \caption{Interview 3, part 1: indicators, questions and follow-up/optional questions.}
    \label{tab:interview_3_questions_part_1}
\end{table}

\begin{table}[]
    \scriptsize
    \centering
    \begin{tabular}{p{3cm}p{6cm}p{5cm}}
        \toprule
        Topic/Indicator & Question & Follow-up/optional question \\
        \midrule
        \rowcolor{teal!10}
        \textbf{Dataset composition} && \\
        11. Instances & - What do the instances that comprise the dataset represent? & \\
        & - How many instances are there in total? & \\
        & - Does the dataset contain all possible instances or is it a sample of instances from a larger set? & \\
        & - What data does each instance consist of? & \\
        & - Is there a label or target associated with each instance? & \\
        & - Is any information missing from individual instances? & \\
        & - Are relationships between individual instances made explicit? & \\
        12. Splits & - How did you split the data for your project? & \\
        & - Do you recommend the same/different splits for other projects? & \\
        13. Representation & - Are there any errors, sources of noise or redundancies in the dataset? & \\
        & - Is the dataset self-contained or does it link to or otherwise rely on external resources? & \\
        \midrule
        \rowcolor{teal!10}
        \textbf{Dataset Processing} & & \\
        14. Preprocessing & - Was any preprocessing or cleaning or labeling of the data done? & \\
        & - Is the software that was used to preprocess or clean or label the data available? & \\
        \midrule
        \rowcolor{teal!10}
        \textbf{Dataset Maintenance} && \\
        15. {Updates} & - Will the dataset be updated? \\
        16. {Storage} & - If the dataset relates to people, are there applicable limits on the retention of the data associated with the instances? & \\
        & - Will older versions of the dataset continue to be hosted and maintained? & \\
        & - How do you store your data? & \\
        \midrule
        \rowcolor{teal!10}
        \textbf{Ethical aspects} && \\
        17. {Review processes and impact assessment} & - Were any ethical review processes conducted? & \\
        & - Is the data minimization principle being applied? & \\
        & - If the dataset relates to people, has an analysis of the potential impact of the datasets and its use on data subjects been conducted? & \\
        18. {Notification and consent duties} & - If the dataset relates to people, were the individuals in question notified about the data collection?	& \\
        & - If the dataset relates to people, did the individuals in question consent to the collection and use of their data? & \\
	    && - If consent was obtained, were the consenting individuals provided with a mechanism to revoke their consent in the future or for certain uses? \\
        19. {Sensitive information} & - Does the dataset contain data that might be considered confidential, offensive or insulting, or sensitive? & \\
        & - If the dataset relates to people, does the dataset identify any subpopulations?	& \\
        & - If the data relates to people, are they directly or indirectly identifiable from the data?	& \\
	    && - If the data relates to people, did you anonymize it? \\
        \midrule
        \rowcolor{teal!10}
        \textbf{AI system reflection} && \\
        20. Reflection on alternatives & - Is there anything that you would do differently about the AI system, if you developed it again? & \\
        \bottomrule
    \end{tabular}
    \caption{Interview 3, part 2: indicators, questions and follow-up/optional questions.}
    \label{tab:interview_3_questions_part_2}
\end{table}

\begin{table}[]
    \scriptsize
    \centering
    \begin{tabular}{p{3cm}p{6cm}p{5cm}}
        \toprule
        Topic/Indicator & Question & Follow-up/optional question \\
        \midrule
        \rowcolor{teal!10}
        \textbf{Social sustainability} && \\
        1. Transparency & - Do you use any method to increase the transparency and explainability of your model? If yes, please describe them. & \\
        & - How do you document your AI system and who can access this documentation? & \\
        & - How do you document the output of your model (audit trails) and who can access this documentation? & \\
        2. Non-discrimination & - Has your application a potential of discrimination. If yes, did you (technically) assess it? & \\
        && - Are processes in place to test data against biases? \\
        && - Have you undertaken an analysis of the data to prevent societal and historical biases in data? \\
        && - Is the data well-balanced and does it reflect the diversity of the targeted end-user population? \\
        && - Are there any differences you can foresee between the data used for training and the data processed by the AI system which could result in the AI system producing discriminatory outcomes or performing differentially for different groups? \\
        && - Have you developed a process to document how data quality issues can be resolved during the design process? \\
        && - Did you put in place educational and awareness initiatives to help AI designers and developers gain awareness of the possible bias they can introduce through the design and development of the AI system? \\
        & - Which methods or metrics did you rely upon? & \\
	    && - Which type of a metric did you choose and why: 1) group fairness metric; 2) individual fairness metric; 3) counterfactual fairness? \\
        & - Did you deploy any methods to eliminate discrimination? If yes, which ones? \\	
        3. Privacy & - Do you use any method regarding the privacy of individuals that use the AI system? & \\
	    && - Did you deploy any methods that improve the privacy of the model? \\
	    && - Is privacy by design being applied in the system? Please elaborate how. \\
        4. Safety and security & - Do you use any method regarding the safety / security of your AI system (system manipulation, data poisoning/corruption, data safety, cybersecurity)? & \\
	    && - What measures were put in place to ensure the safety and security of the AI system and protect it from system manipulation, data poisoning or corruption? \\
	    && - What measures were put in place to ensure the safety and security of the data processed by the AI system? \\
        5. Additional harms & - Are you aware of any additional harms and do you engage in their mitigation? & \\
	    && - Generative AI: Did you consider hallucination and fake or misleading content that could be generated by the AI system? \\
	    && - Generative AI: Did you consider possible IP rights infringement by the AI system? \\
        \midrule
        \rowcolor{teal!10}
        \textbf{Environmental sustainability} && \\
        6. {Hardware overview} & - Describe your hardware infrastructure for the project. & \\
        7. Efficiency of AI system & - Have you considered the efficiency of your AI system? If yes, how? & \\
	    & - Did you take any measures to increase efficiency of your AI system such as: 1) the selection of low complexity models, 2) using pretrained models and transfer learning, 3) using model compression methods, 4) using methods for efficient model training, 5) using methods to reduce the amount of data? & \\
	    & - Do you measure the model efficiency? If yes, how? & \\
        && - Based on your considerations or measuring: Which part of the AI life-cycle is most critical? \\
        8. Energy and emissions & - Do you consider the carbon footprint and energy consumption of your model, or do you track it? & \\
	    & - If you track it, during which stages do you track it and how? & \\
		&& - Based on your considerations or measuring: Which part of the AI life-cycle is most critical? \\
        \bottomrule
    \end{tabular}
    \caption{Interview 4, part 1: indicators, questions and follow-up questions.}
    \label{tab:interview_4_questions_part_1}
\end{table}

\begin{table}[]
    \scriptsize
    \centering
    \begin{tabular}{p{3cm}p{6cm}p{5cm}}
        \toprule
        Topic/Indicator & Question & Follow-up/optional question \\
        \midrule
        9. Materials & - Have you reflected on the material used for your project? & \\
        & - Which resources do you see as critical for AI development in general?	& \\
	    & - What are your plans about your system, once it is decommissioned? & \\
        10. Certificates & - Do you use certified hardware or did you consider the use? & \\	
	    & - Do you rely on certified data centers or consider migrating to one? & \\
		&& - What type of certificates are they? \\
        \midrule
        \rowcolor{teal!10}
        \textbf{Economic sustainability} && \\
        11. \textbf{Model sharing} && \\
        11.1 Sharing & - How will the model be shared? & \\
        & - When will the model be shared? & \\
        & - What are the citation details of the model? & \\
        & - What is the license of the model? & \\
        & - Who is the contact person for the model and how can that person be reached? & \\
        & - Who will be hosting and maintaining the model? & \\
        11.2 Use & - What (other) tasks could the model be used for? & \\
        & - What are the out-of-scope applications of your model? & \\
        11.3 Compatibility & - With which datasets, libraries, or environments is your model compatible with? & \\
        12. \textbf{Dataset sharing} && \\
        12.1 Sharing & - How will the dataset be shared? & \\
	    & - When will the dataset be shared? & \\
	    & - What are the citation details of the data? & \\
        & - What is the license of the data? & \\
	    & - Who is the contact person for the dataset and how can that person be reached? & \\
	    & - Who will be hosting and maintaining the dataset? & \\
        12.2 Use & - What (other) tasks or contexts could the dataset be used for? & \\
	    & - Are there tasks or contexts for which the datasets should not be used? & \\
        12.3 Compatibility & - With which models/libraries/environments is your dataset compatible with? & \\
        \bottomrule
    \end{tabular}
    \caption{Interview 4, part 2: indicators, questions and follow-up/optional questions.}
    \label{tab:interview_4_questions_part_2}
\end{table}

\end{document}